\documentclass{aa}

\usepackage{graphicx}
\usepackage{txfonts}
\begin{document}

   \title{Formation, diffusion and accreting pollution of DB white dwarfs}


   \author{Chunhua Zhu
          \inst{1,2}
          \and
          Helei Liu\inst{1,2}
          \and
          Zhaojun Wang\inst{1,2}
          \and
          Guoliang L\"{u}\inst{1,2}
          }

   \institute{School of Physical Science and Technology,
               Xinjiang University, Urumqi, 830046, China\\
              \email{chunhuazhu@sina.cn}
         \and
             Center for Theoretical Physics, Xinjiang University, Urumqi, 830046, China\\
             }



  \abstract
   {Over 1500 DBZ or DZ white dwarfs (WDs) have been observed so far, and polluted atmospheres
    with metal elements have been found among these WDs. The surface heavy element abundances of known DBZ or DZ WDs show an evolutionary sequence.
   The cooling, diffusion and accretion are important physical processes in the WD evolution which can alter the
   element abundances of the WD surface. }
   {Using the stellar evolutionary code, we investigate the DB WD formation and the effects of input parameters$-$mixing length parameter ($\alpha_{\rm MLT}$),
    thermohaline mixing efficiency ($\alpha_{\rm th}$) and the metallicity ($Z$)$-$on the structures of these DB WDs.
    The impacts of convective zone mass ($M_{\rm cvz}$),
   cooling timescales,  diffusive timescales ($\tau_{\rm diff}$),
   and mass-accretion rate ($\dot{M}_{\rm a}$) on the element abundances of the WDs' surfaces are discussed.
   By comparing the theoretical model results with observations, we try to understand the evolutionary sequence of the heavy element abundance on DBZ WD surfaces.}
   {By using Modules for Experiments in Stellar Evolution, we create DB WDs, and
   simulate the element diffusion due to high gravitational fields and the
   metal-rich material accretion coming from the planet disrupted by the WD. Then,
    we calculate the element abundances of these DB WDs for the further comparison with observations.   }
   {In our models, the input parameters ($\alpha_{\rm MLT}$, $\alpha_{\rm th}$ and $Z$) have very weak effect on DB WD structures
    including interior temperatures, chemical profiles and convective zones. They hardly affect the evolution of
    the heavy elements on the surface of DB WDs. The mass-accretion rate and the effective temperature of
    DB WDs  determine the abundances of heavy elements. The evolutionary sequence
    of Ca element for about 1500 observed DB or DBZ WDs cannot be explained by the model with a constant mass-accretion rate,
    but is consistent well with the model in which the mass-accretion rate decreases by one power law when $T_{\rm eff}>10$ kK
    and  slightly increases by another power law when $T_{\rm eff}<10$ kK. }
   { The observed DB WD evolutionary sequence of heavy element abundances originates from WD cooling
   and the change of mass-accretion rate. }


   \maketitle
%

\section{Introduction}
It is well known that single stars with initial mass between $\sim1$ and 8 M$_\odot$
finally evolve into white dwarfs (WDs). Due to high gravitational fields ($\log \ g \sim$8 cm s$^{-2}$),
the heavy elements on WDs' surfaces would diffuse downward during WD cooling.
Usually, the timescale of diffusion ($\tau_{\rm diff}$) at the photosphere is about several to 10${^4}$ years\citep{Koester2009},
which is much shorter than cooling timescale ($t_{\rm cool}$) ($\sim 10^9$ yr)\citep[e. g.,][]{Shapiro1983,Zhu2019,Lu2017,Lu2020}.
Therefore, cool WDs should have pure hydrogen (H) or helium (He) atmospheres. The former
is called as DA WD, while the latter is called as DB WD. However, \cite{Zuckerman2003}
pointed out that more than 25\% of WDs are polluted by metal elements such as Mg, Fe, Na.
WDs are called as DAZ or DBZ WDs if their spectra show H or He lines with heavy element lines.
When only heavy elements lines are displayed in the spectra, WDs are categorized as DZ type.
Three different ways are proposed as three possible sources for the surface heavy elements of the WDs,
such as primordial or fallback stellar material,
interstellar medium, or debris disk produced by
WD tidally disrupting rocky objects (i.e. planets)\citep[e. g.,][]{Farihi2016}.

The pollution of WDs has been explained by ongoing accretion of
planetary debris.
A number of observational evidences show the infrared emission from debris
disk around polluted WDs\citep{Jura2003,Farihi2009,Girven2012,Vanderburg2015},
thus these polluted WDs become unique laboratories for studying
the interior composition of exoplanets\citep{Zuckerman2007,Koester2014,Jura2014}.
The chemical abundances detected on the surface of
polluted WDs reflect the equilibrium between accretion for metal-rich material
and diffusive sedimentation\citep{Koester2009,Bauer2018}.

\cite{Dupuis1992} firstly explored the metal traces in WDs,
and they also investigated the diffusion of metals accreted onto WDs\citep{Dupuis1993a,Dupuis1993b}.
\cite{Koester2006} calculated the diffusion timescales for some metals in DAZ WDs' atmosphere,
and estimated the accretion rates for 38 DAZ WDs.  \cite{Koester2009} extended the
above works to DAZ, DBZ and DZ WDs. Considering diffusion and thermohaline mixing,
\cite{Wachlin2017} and \cite{Bauer2018} simulated the trace of metals for DAZ WDs. \cite{Bauer2019} discussed
the effects of the mixing processes (including convection, gravitational sedimentation, overshoot,
and thermohaline instability) on the diffusion. Using new WD envelope models and diffusion,
\cite{Koester2020} investigated the atmospheres of carbon-rich WDs.
In the theoretical models, the metal abundances of several polluted WDs can be explained well if
 suitable accretion rates are assumed\citep[e. g.,][]{Koester2009,Bauer2018}. They
 also predicted that the metals would rapidly settle downward as soon as accretion stops.
 However, as shown in \cite{Koester2009}  and \cite{Bauer2018}, the diffusion timescales of
 WDs increase with their effective temperature ($T_{\rm eff}$) decreasing. For a WD with $T_{\rm eff}<10$ kK,
 $\tau_{\rm dif}$ is longer than about $10^6$ yr.
 Therefore it is imperfect to check long-timescale diffusion theory by only comparing theoretical results with
 several known cool WDs. A comparison involved a large observational sample of DB WDs with different  $T_{\rm eff}$s
 becomes necessary.

Thanks to many large sky surveys, the number of observed WDs
are dramatically increasing\citep[e. g.,][]{Gaia2016,Gaia2018,Chambers2016,Blouin2019}.
Up to now, there are more than 60000 WDs in \emph{The Montreal White Dwarf Database}\citep{Dufour2017}.
One thousand and twenty three of them are DBZ or DZ WDs\citep{Coutu2019}.
Observationally, \cite{Dufour2007} showed the  spectroscopic and photometric data of 147 DZ WDs with
$T_{\rm eff}$ between about 6 kK and 12 kK. Based on SDSS DR10 and 12, \cite{Koester2015}
analyzed the data of 1107 DBZ WDs whose effective temperatures are between about 50 kK and 11 kK.
\cite{Hollands2017} identified 231 cool DZ WDs with $T_{\rm eff}$ lower than 9 kK in SDSS DR12.
They discussed the distribution of log[Ca/He] vs. $T_{\rm eff}$ for the three samples (See Figure 11 of
\cite{Hollands2017}).
At about $T_{\rm eff}$ > 10 kK,  Ca abundances rapidly decrease with $T_{\rm eff}$ declining.
\cite{Koester2015} suggested that this trend should be relative to the mass-accretion rates.
However, Ca abundances of DZ WDs with $T_{\rm eff}$ between 10 kK and 8 kK increase by
about 100 times. \cite{Hollands2017} considered that this sharp increase might result from
the decrease of convective zone mass ($M_{\rm cvz}$) or the increase of $\tau_{\rm diff}$.
The second downwards trend of Ca abundance with $T_{\rm eff}$ appears between 9 kK and
4 kK. \cite{Hollands2018} suggested that the trend is relative to  $M_{\rm cvz}$ or $\tau_{\rm diff}$.

Compared with DAZ WDs whose metal pollution is monotonically decreasing with $T_{\rm eff}$ declining\citep{Koester2014},
DBZ WDs have more complicated progresses for metal pollution.
The main reason is that DB WDs undergo different formation channels, and have distinctive interior structures.
In this work, employing the stellar evolution code, we investigate the physical mechanisms to explain the surface metal abundance of the
polluted DB WD. The model descriptions are
given in section 2. DB WD's properties and their accretion pollution are shown and discussed in sections 3 and 4.
The paper is closed with conclusions in section 5.

\section{Models}
In the present paper, we use \emph{Modules for Experiments in Stellar Evolution}
(MESA, [rev. 12115]; \cite{Paxton2011,Paxton2013,Paxton2015,Paxton2018,Paxton2019})
to create He-rich WDs without H which are noted as DB WDs, simulate the element diffusion within them
and metal-rich material accretion. There are many factors to change the element abundances on the
WD surface. \cite{Bauer2019} discussed the effects of
convection, thermohaline instability, gravitational diffusion and rotation on the element mixing of WDs.
Because the rotations velocities of most isolated WDs observationally are low\citep{Berger2005,Kawaler2015,Hermes2017}, we do not consider the rotation.
Convection directly determines the timescale of element diffusion\citep{Koester2009,Bauer2018}.
In the present paper, we adopt the ML2 convection prescription\citep{Bohm1971,Tassoul1990},
and use Ledoux criterion for convection.
The table named as 'DB$\_$WD$\_$tau$\_$25' in MESA ( It is helium dominated atmosphere table for DB WDs)
is used to calculate the DB WD atmosphere boundary.
The size of convective zone depends on mixing length parameter ($\alpha_{\rm MLT}$).
In order to discuss its effect, we take $\alpha_{\rm MLT}$ as 0.8 and 1.8 in different simulations, respectively.

\cite{Deal2013} and \cite{Wachlin2017} considered that thermohaline mixing
can change the element abundances on the surfaces of polluted WDs.
MESA adopts the method of \cite{Kippenhahn1980} to calculate the effects of thermohaline
mixing, in which parameter $\alpha_{\rm th}$ is used to give mixing efficiency.
In our work, $\alpha_{\rm th}$ is taken as 0, 1 and 1000 in different calculations for
testing its effects.

\cite{Schatzman1945} suggested that the high gravitational fields in cool WDs
should result in the downward diffusion of heavy elements.
By resolving the Burgers equations which give multicomponent fluid's evolutions\citep{Burgers1969},
\cite{Thoul1994} investigated the element diffusion in the interior of the Sun.
Using the approach in \cite{Thoul1994},
MESA can calculate the chemical diffusion in stellar interior\citep{Paxton2015,Paxton2018}.
The diffusion coefficients originated from \cite{Paquette1986} and updated by \cite{Stanton2016} are
used in our models.

Similarly, metallicity ($Z$) can also affect stellar evolutions and WD properties.
Here, Table \ref{tab:case} gives all cases in which different input parameters are considered.

\begin{table}
  \caption{All cases in the present paper are simulated. The first column gives
           the case number. Columns 2, 3 and 4 show the values of input parameters
            $\alpha_{\rm MLT}$, $\alpha_{\rm th}$ and $Z$, respectively.
           }
  \begin{tabular}{cccc}
  \hline
   Cases & $\alpha_{\rm MLT}$ &$\alpha_{\rm th}$&$Z$\\
   case 1 & 1.8 & 1 & 0.02 \\
   case 2 & 0.8 & 1 &0.02\\
   case 3 & 1.8 &0  & 0.02 \\
   case 4 & 1.8 & 1000& 0.02\\
   case 5 & 0.8 & 1   &0.001\\
   \hline
 \label{tab:case}
\end{tabular}
\end{table}

\section{Formation and structures of DB WDs}
Many observations have showed that there are some H elements in the atmospheres of DBZ or DZ WDs
\citep{Voss2007,Koester2015,Coutu2019}. However, the ratios of H to He abundance estimated by these observations
are lower than about $10^{-2}$.
These H elements maybe continuously be accreted by DBZ or DZ WDs from interstellar medium\citep{Voss2007,Koester2015}.
Therefore, there may be no H elements left in the atmospheres of DB WDs when they form.

\begin{figure}
\includegraphics[totalheight=3.0in,width=3.0in,angle=-90]{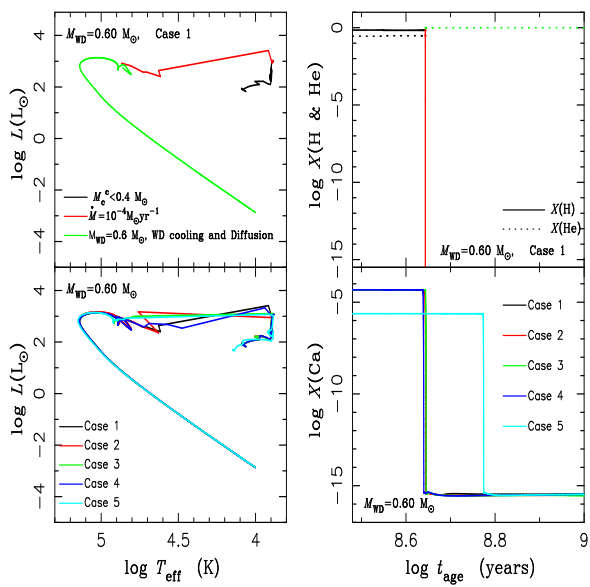}
\caption{The 0.6 M$_\odot$ DB WD produced by main sequence star with an initial mass of 3.5 M$_\odot$.
         The left-top panel gives the evolution in HR diagram for the star in case 1 ($\alpha_{\rm MLT}=$1.8,
         $\alpha_{\rm th}=1$ and $Z=0.02$), in which the lines with different colors represent different evolutionary phases.
         The left-bottom panel shows the evolutions in HR diagram for stars with different input parameters which
         are given by different colors. The right-top panel is similar with the left-top panel, but for evolution of H and He abundances
         on the stellar surface. The right-bottom panel is similar with the left-bottom panel, but for evolution of Ca abundance
         on the stellar surfaces. The details can be seen in the text.
         }
\label{fig:hr0.6}
\end{figure}

Usually, the range of DBZ or DZ WDs' masses is between about 0.4 and 1.0M$_\odot$ and
their mass distribution has a peak around 0.6 M$_\odot$\citep[e. g.,][]{Han1998,Han2000,Coutu2019}.
Take 0.6 M$_\odot$ DB WD created by main sequence (MS) star under input parameters in case 1 as an example,
we give all details for creating DB WDs by
the following steps:\\
(i)The first step is showed by the black line in the left-top panel of Figure \ref{fig:hr0.6}.
The 3.5 M$_\odot$ MS star begins to normally evolve, that is, H starts to burn in the stellar core.
The mass-loss rate ($\dot{M}$) is calculated by 'Dutch' scheme\citep{Paxton2011}, in which $\dot{M}$ of hot and cool stars
is given by \cite{Nieuwenhuijzen1990,Nugis2000,Vink2001,Glebbeek2009} and \cite{Reimers1975}, respectively.
The element mixing is mainly determined by convection and thermohaline instability. At this phase,
in order to save CPU time, we do not consider gravitational diffusion. \\
(ii)The second step is showed by the red line in the left-top panel of Figure \ref{fig:hr0.6}.
We artificially enhance the mass-loss rate up to $10^{-4}$M$_\odot$ yr$^{-1}$ when He-core mass is
larger than 0.6 M$_\odot$. The H-rich envelope is rapidly stripped, and the star evolves into He star.
As the red lines in the right-top panel of Figure \ref{fig:hr0.6} shows, the H abundance ($X({\rm H})$) on the stellar surface
decreases from about 0.7 to about lower than $10^{-15}$, while $X({\rm He})$ increases up to about 0.98.\\
(iii)The third step is WD cooling, which is given by the green line. At this time, all H element almost is lost. He element
is lightest, and it floats upward stellar surface by gravitational settling. A DB WD is created.

The left-bottom panel of Figure \ref{fig:hr0.6} shows the evolution in HR diagram for the star
with different $\alpha_{\rm MLT}$, $\alpha_{\rm th}$ and $Z$. Obviously, the effects of input parameters on evolutionary tracks
are negligible. The right-bottom panel gives the change of $X({\rm Ca})$ on the stellar surface.
$X({\rm Ca})$ on the stellar surface starts to reduce because of gravitational sedimentation at WD cooling phase.

\begin{figure}
\includegraphics[totalheight=3.0in,width=3.0in,angle=-90]{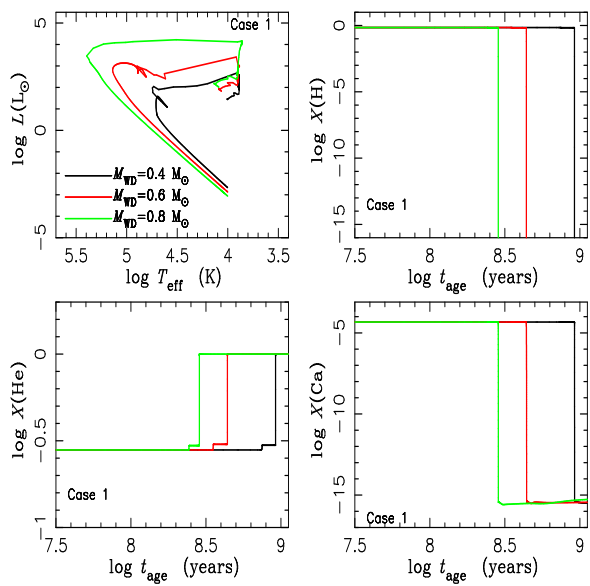}
\caption{Similar with Figure \ref{fig:hr0.6}, but for 0.4, 0.6 and 0.8 M$_\odot$
         DB WDs produced by MS stars in case 1 with initial masses of 2.5, 3.5 and 5 M$_\odot$,
         which are represented by black, red and green lines, respectively.}
\label{fig:hr}
\end{figure}

Using similar method, we also create DB WDs with 0.4 and 0.8 M$_\odot$, which
are showed in Figure \ref{fig:hr}. The changes
of $X({\rm H})$, $X({\rm He})$ and $X({\rm Ca})$ on these WD surfaces are given, too.

\begin{figure*}
\begin{tabular}{ccc}
\includegraphics[totalheight=3.0in,width=3.0in,angle=-90]{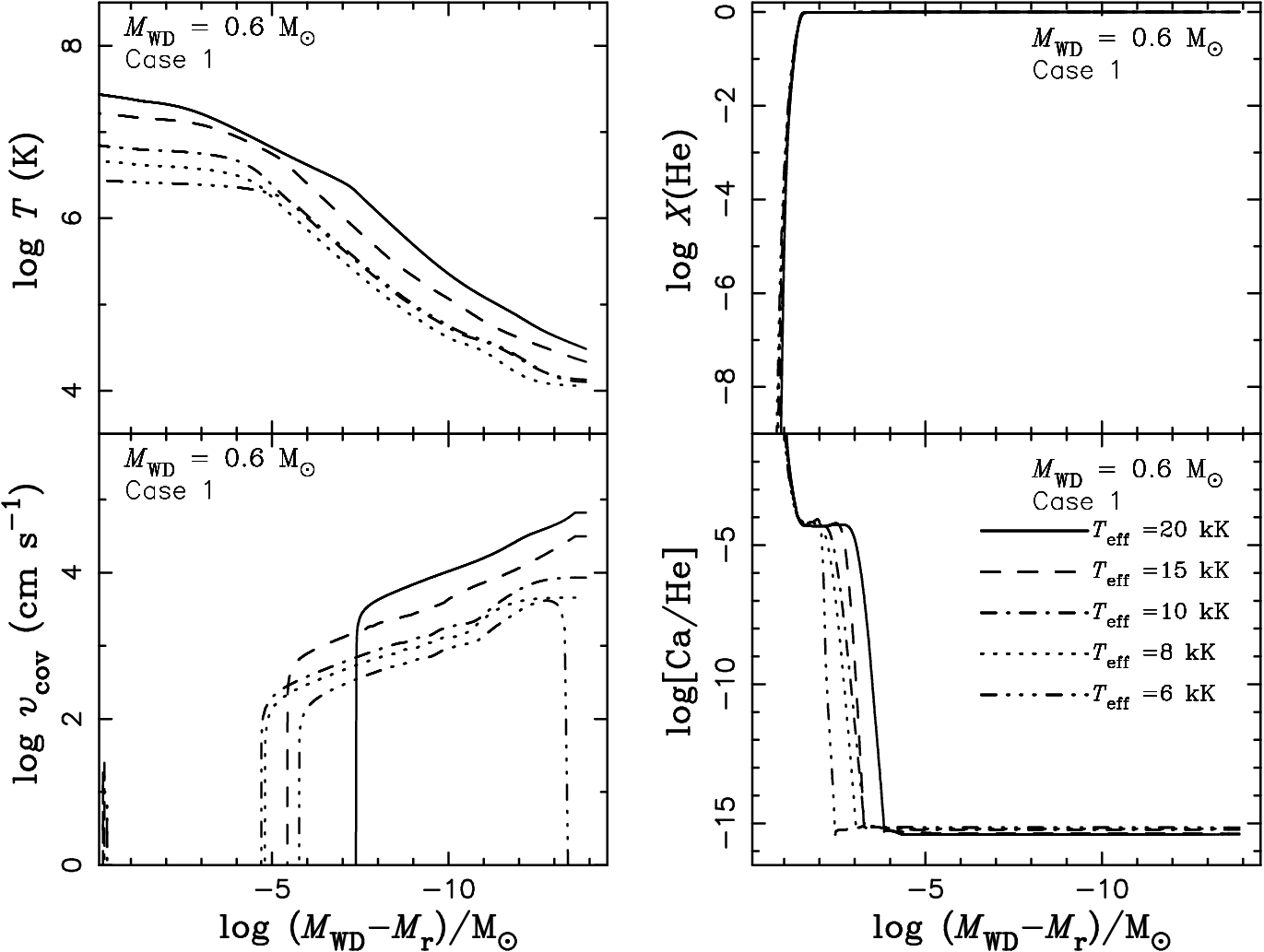}&
\includegraphics[totalheight=3.0in,width=3.0in,angle=-90]{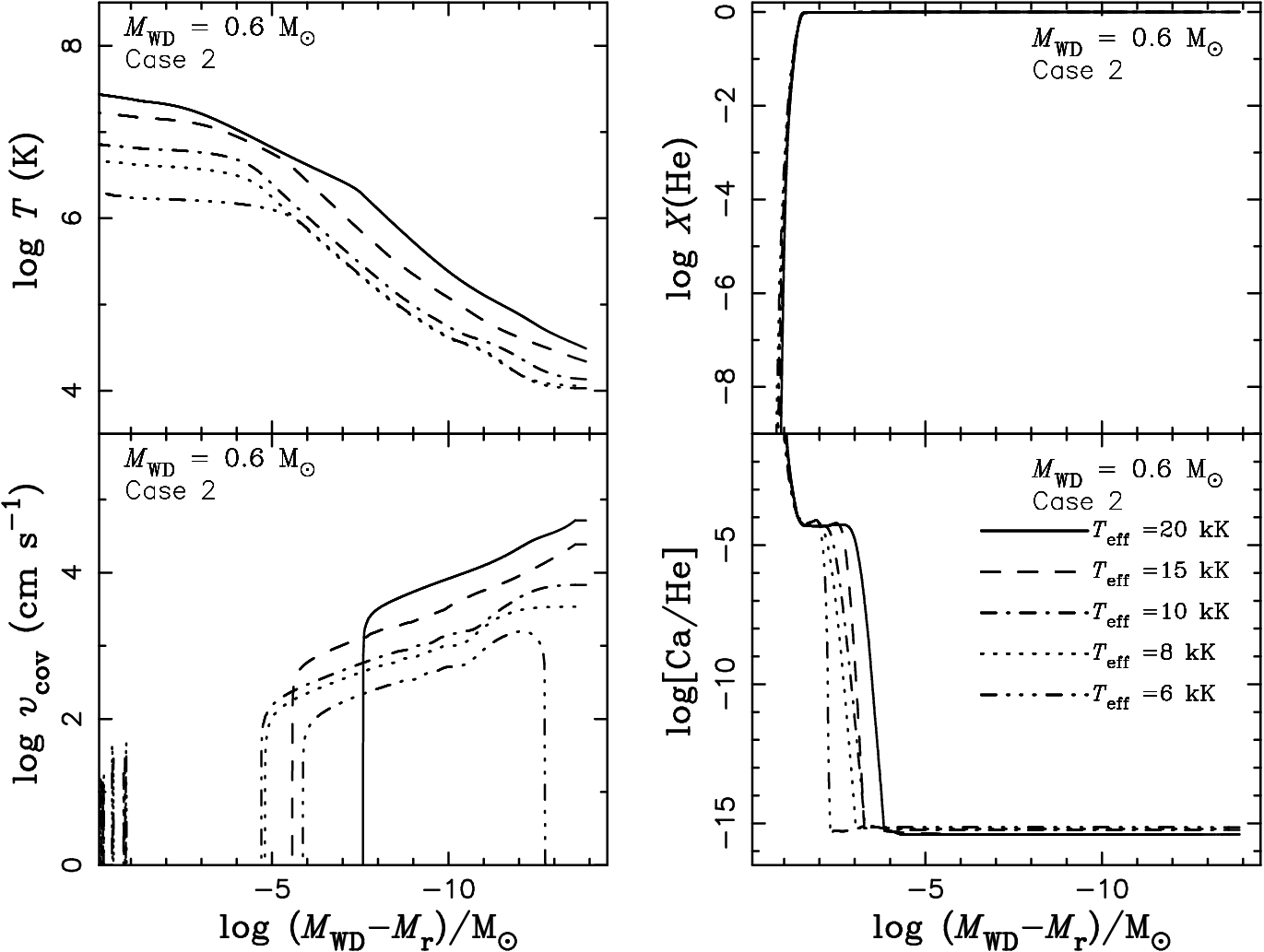}&\\
\includegraphics[totalheight=3.0in,width=3.0in,angle=-90]{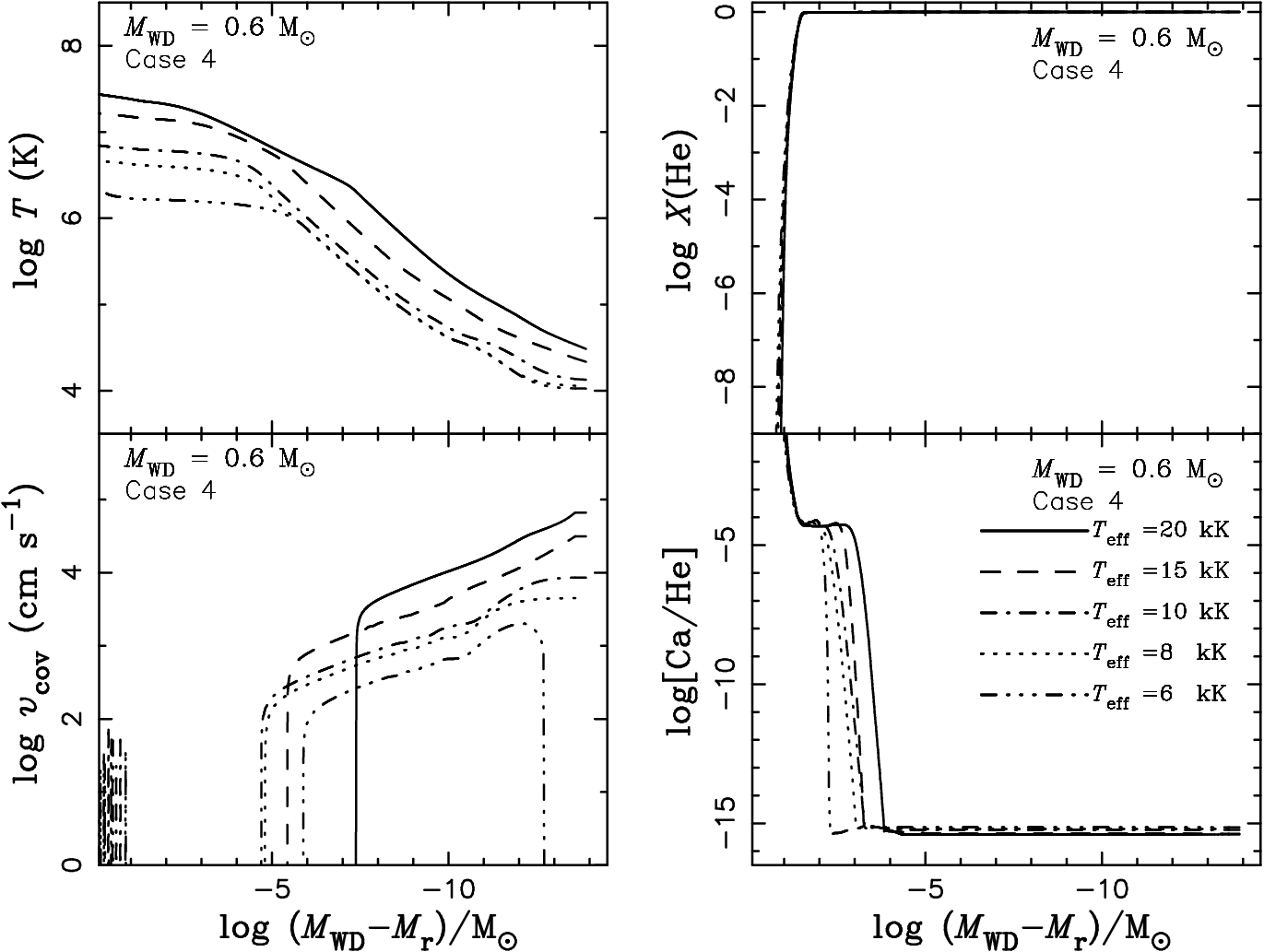}&
\includegraphics[totalheight=3.0in,width=3.0in,angle=-90]{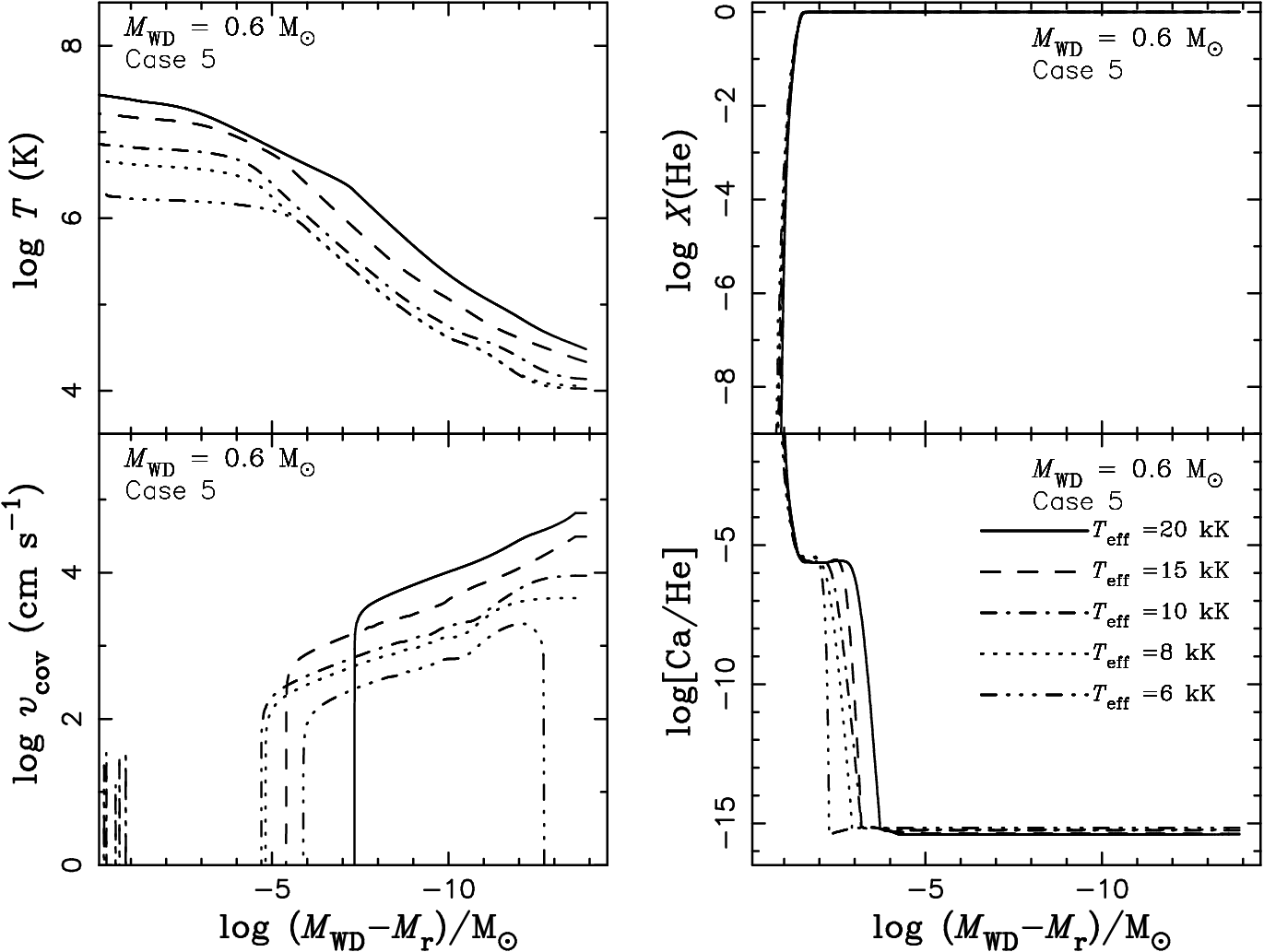}\\
\end{tabular}
\caption{Structures of 0.6 M$_{\odot}$ DB WDs in different effective temperatures ($T_{\rm eff}$s)
        for cases 1, 2, 4 and 5, respectively. The different $T_{\rm eff}$s are given by different lines.
        $T$, $v_{\rm cov}$, $X{\rm (He)}$ and [Ca/He]
        represent the temperature, convective velocity, He abundance and the  abundance ratio of Ca to He,
        respectively.  }
\label{fig:stru}
\end{figure*}

As the left-bottom panel of Figure \ref{fig:hr0.6} shows, the cooling tracks of
DB WDs are hardly affected by the input parameters.
Similarly, the effects of these input parameters on DB WD internal structures
can be negligible.

\begin{figure*}
\begin{tabular}{cc}
\includegraphics[totalheight=3.2in,width=3.5in,angle=-90]{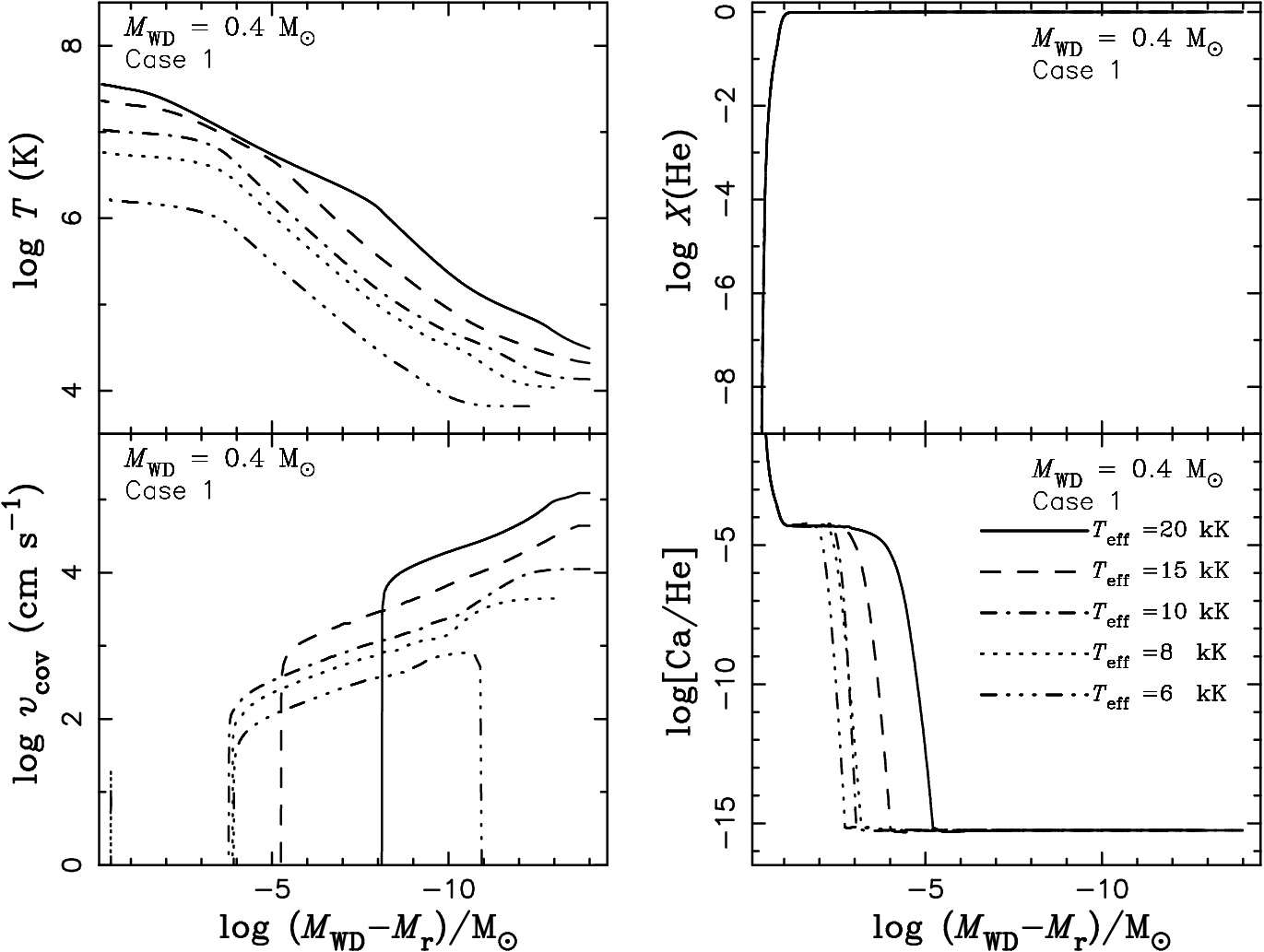}&
\includegraphics[totalheight=3.2in,width=3.5in,angle=-90]{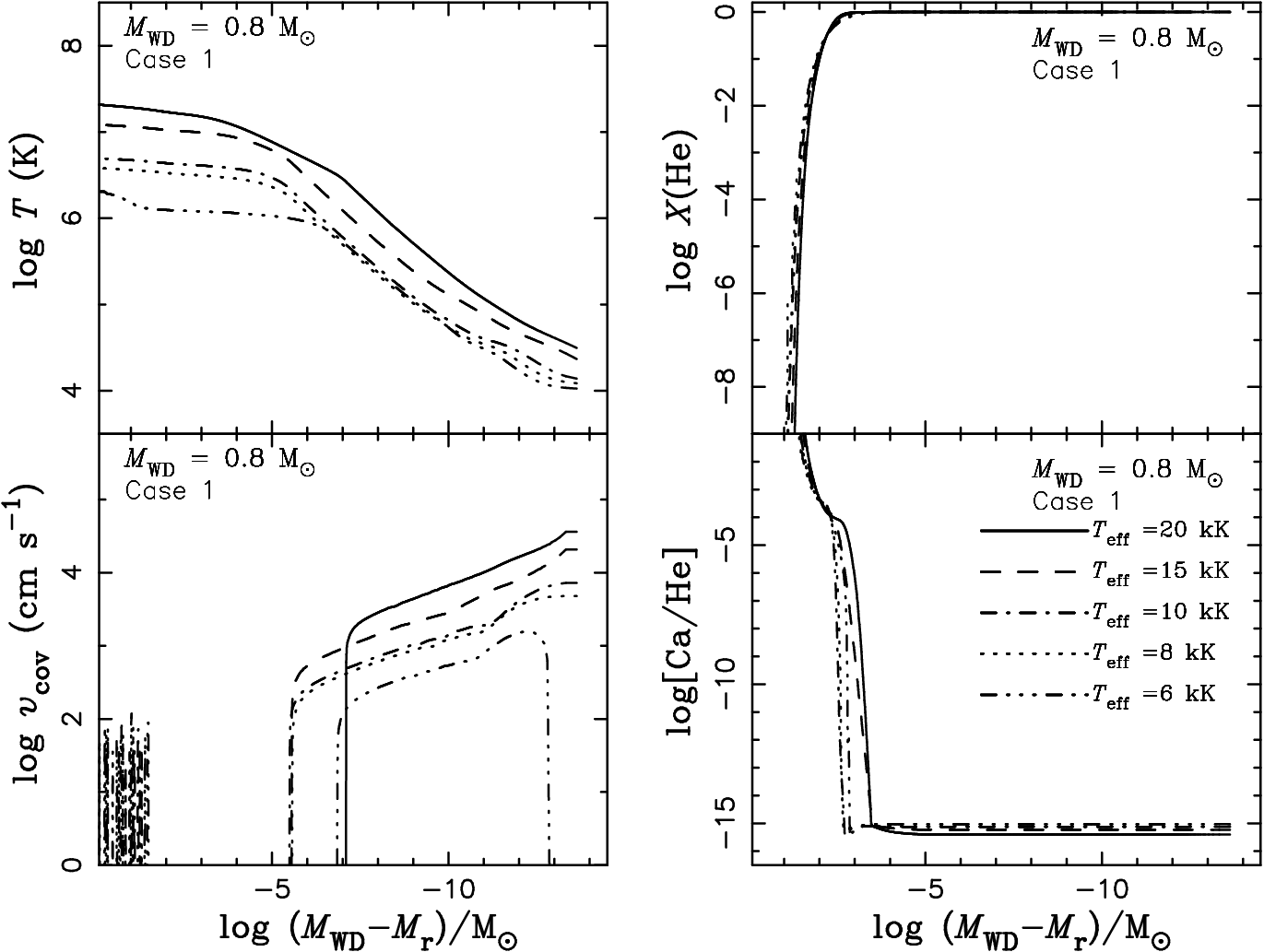}\\
\end{tabular}
\caption{Similar with Figure \ref{fig:stru}, but for 0.4 and 0.8 M$_\odot$ DB WDs for case 1.
       }
\label{fig:stru2}
\end{figure*}

In Figure \ref{fig:stru}, we find that the profiles of
the temperature, convective velocity ($v_{\rm cov}$),
He abundance ($X{\rm(He)}$) and the abundance ratio of Ca to He ([Ca/He])
for 0.6 M$_\odot$ DB WD at the same effective temperature are similar.
Due to the strong gravitational diffusion of WD, heavy elements sink down and light He element
floats up. For example, $X{\rm(Ca)}$ on the DB WD's surface has decreased to
$10^{-15}$ from initial $10^{-5}$, while a heavy He envelope with mass of about $0.02$ M$_\odot$
forms around WD surface.
Figure \ref{fig:stru2} gives the profiles of 0.4 and 0.8 M$_\odot$ DB WDs for case 1.
Obviously, in our model, He layer mass is affected by WD's mass.
It changes from about 0.1 to 0.01 M$_\odot$ when $M_{\rm WD}$ increases from
0.4 to 0.8 M$_\odot$.

\begin{figure}
\includegraphics[totalheight=3.0in,width=3.0in,angle=-90]{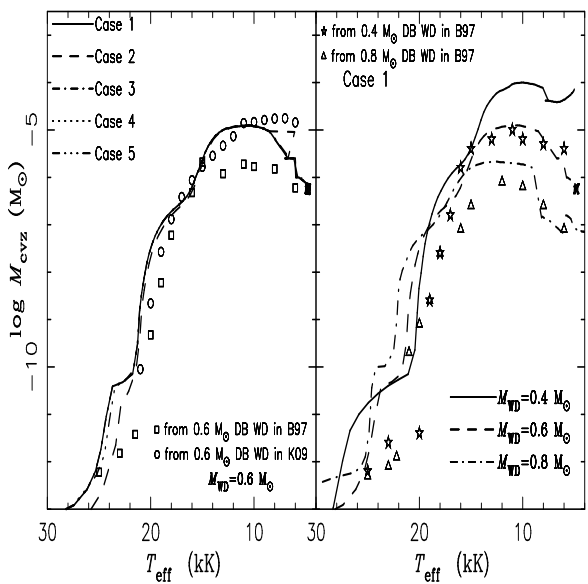}
\caption{The mass of convective zone ($M_{\rm cvz}$) vs. the WD's effective temperature ($T_{\rm eff}$).
         The left panel is for 0.6 M$_\odot$ DB WD in different cases, while right panel is
         for 0.4, 0.6 and 0.8 M$_\odot$ DB WDs in case 1. Theoretical results from \cite{Benvenuto1997} and \cite{Koester2009}
         are showed by different symbols.
         B97 and K09 refer to \cite{Benvenuto1997} and \cite{Koester2009}, respectively.
          }
\label{fig:sconvt}
\end{figure}

In Figure \ref{fig:sconvt}, we give the change of convective-zone mass ($M_{\rm cvz}$) around WD surface with $T_{\rm eff}$.
For 0.6 M$_\odot$ DB WD showed in the left panel of Figure \ref{fig:sconvt},
the effects of input parameters on $M_{\rm cvz}$ can be negligible.
The main reasons are as follows:\\
(i)The mixing length parameter $\alpha_{\rm MLT}$ has a weak effect on $M_{\rm cvz}$
because high density of WDs results a small pressure scale height. For example, it
is about 10 cm for a WD with $T_{\rm eff}=6$ kK.\\
(ii)The thermohaline mixing hardly affects the convective zone of DB WDs, while it can significantly
affect $M_{\rm cvz}$ of DA WDs\citep{Wachlin2017,Bauer2018}.
Compared with the latter ($10^{-15}-10^{-11}$ M$_\odot$ when $T_{\rm eff}>$ 10 kK)\citep{Koester2009,Wachlin2017},
$M_{\rm cvz}$ of DB is much massive, and between about $10^{-9}-10^{-5}$ M$_\odot$.
Thick convective zone of DB WDs dilutes the effects of thermohaline mixing, which has been discussed by \cite{Bauer2019}.
Simultaneously, \cite{Bauer2019} mentioned that the mean molecular weight of DB WD is more than two times of DA WD, which dilutes
thermohaline mixing effects. \\
(iii)Metallicity has no effect on $M_{\rm cvz}$ because the heavy elements rapidly diffuse downward due to
the strong gravitational field of WDs.

Compared with $M_{\rm cvz}$ of 0.6 M$_\odot$ DB WD calculated by \cite{Benvenuto1997} and \cite{Koester2009},
$M_{\rm cvz}$ in this work is similar with their results when $T_{\rm eff}>\sim$ 14 kK.
$M_{\rm cvz}$ in this
work is between that in \cite{Benvenuto1997} and \cite{Koester2009} when  $T_{\rm eff}<\sim$ 14 kK.
The right panel of
Figure \ref{fig:sconvt} shows $M_{\rm cvz}$  in the models of 0.4 and 0.8 M$_\odot$ DB WDs.
Compared with the results of \cite{Benvenuto1997}, $M_{\rm cvz}$ in this work is more massive.
The differences mainly result from the following possible aspects:\\
Firstly, in \cite{Benvenuto1997}, the He layer mass of DB WDs is between about $10^{-2}$ and $10^{-6}$ M$_\odot$.
However, in our work, we consider the gravitational diffusion in DB WDs.
The He layer mass is larger than $10^{-2}$ M$_\odot$ and the heavy elements (such as Ca, Fe et al.) sink
down. The different chemical profile around WD surface can affect the convective zone.\\
Secondly, in \cite{Benvenuto1997} and \cite{Koester2009}, $M_{\rm cvz}$ is defined by the thermal time scale. However,
$M_{\rm cvz}$ is defined by Ledoux criterion in our results. As discussed in \cite{Koester2009},
$M_{\rm cvz}$ can differ by orders of magnitude because of different definitions. \\


\section{Accreting pollution of DB WDs }
\begin{figure}
\includegraphics[totalheight=3.0in,width=3.0in,angle=-90]{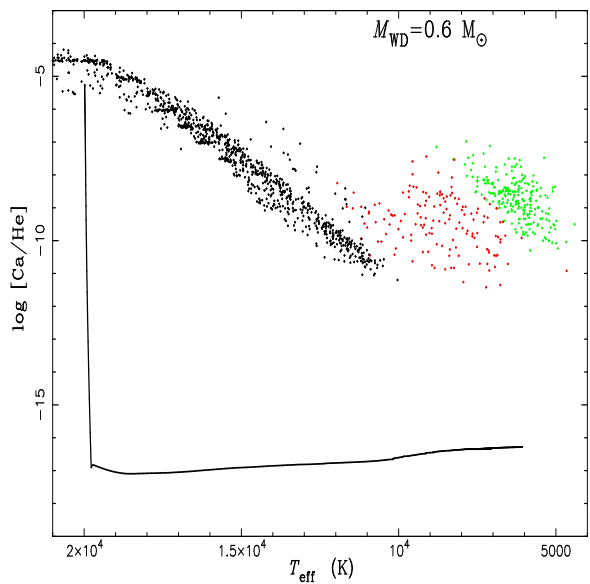}
\caption{The evolutions of log [Ca/He] during 0.6 M$_\odot$ DB WD cooling, in which the gravitational diffusion
is not involved when $T_{\rm}>$20 kK. Black, red and green
dots represent observations from \cite{Koester2015}, \cite{Dufour2007} and \cite{Hollands2017},
respectively.}
\label{fig:chte}
\end{figure}
The right-bottom panels of Figures \ref{fig:stru} and \ref{fig:stru2} show that
[Ca/He] on the surface of DB WD has decreases to about $10^{-15}$ due to gravitational settling
when $T_{\rm eff}>$20 kK.
\cite{Chayer1995a} suggested that some element diffusion can be prevented by radiative levitation
when WD temperature is higher than 20 kK \citep{Chayer1995b,Chayer2014}.
In Figure \ref{fig:chte}, we do a test for 0.6 M$_\odot$ DB WD as follows: The gravitational
settling is not included when $T_{\rm eff}$ of cooling WD is higher than 20 kK,
but it is involved when  $T_{\rm eff}<$ 20 kK. We find that [Ca/He] rapidly decreases,
and can not explain observations.
Therefore, the heavy elements observed on the DB WD's surfaces must originate from other sources.
The rocky objects tidally disrupted by DB WD are possible source\citep{Farihi2016}.

\subsection{Metal-rich Material Accretion}
In general, the element abundances on the surface of accreting DB WD depend on not only the WD properties,
but also  mass-accretion rates ($\dot{M}_{\rm a}$) and the chemical abundances of accreted material.
In order to match the observed properties of G29-38 in \cite{Xu2014}, \cite{Bauer2018} assumed that
the mass fractions of Fe, O, Mg, Si and Ca were in accreted materials 0.307, 0.295, 0.199, 0.153 and 0.046, respectively.
We adopt the above mass fractions.

\begin{figure}
\includegraphics[totalheight=3.5in,width=3.0in,angle=-90]{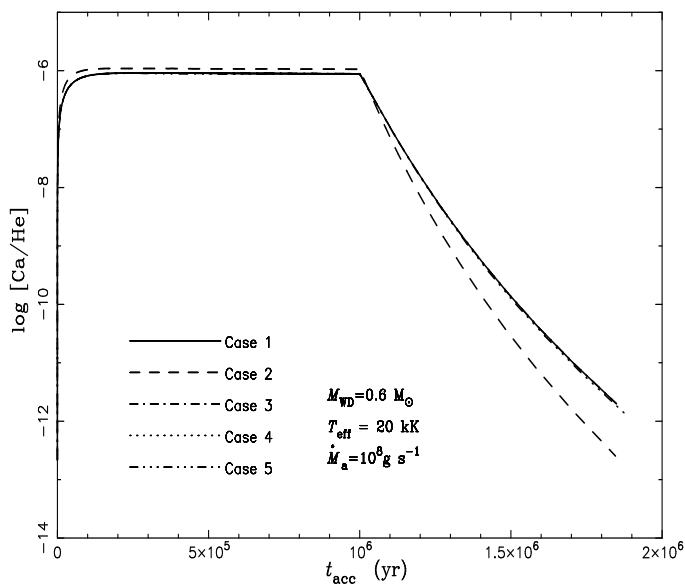}
\caption{The evolution of [Ca/He] on the surface of 0.6 $M_\odot$ DB WD with a mass-accretion rate of $10^8$ g s$^{-1}$
         when $T_{\rm eff}=$ 20 kK. Accretion ceases after $10^6$ years. The different lines represent
         different cases which showed in the left-bottom zone. }
\label{fig:acca}
\end{figure}

By resolving the Burgers equations, MESA can calculate the
chemical diffusion of accreting WD.
Figure \ref{fig:acca} shows the evolution of [Ca/He] on the surface of 0.6M$_\odot$ DB WD
with a mass-accretion rate of $10^8$ g s$^{-1}$ when $T_{\rm eff}=$ 20 kK.
It takes about $10^4$ yr to reach an accretion-diffusion equilibrium for the accreting DB WD.
When the accretion stops, Ca element diffuses downward within a diffusive timescale of about $10^6$ yr,
which is similar with these in \cite{Koester2009}.
Obviously, input parameters ($\alpha_{\rm MLT}$, $\alpha_{\rm th}$ and $Z$) have weak effects on the surface [Ca/He].
The main reasons are similar with these for $M_{\rm cvz}$.

\begin{figure}
\includegraphics[totalheight=3.5in,width=3.0in,angle=-90]{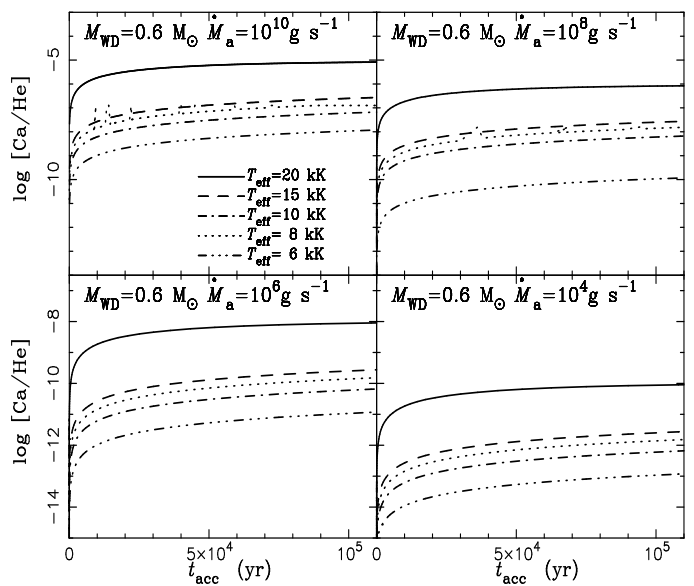}
\caption{The evolution of [Ca/He] on the surface of 0.6 $M_\odot$ DB WD with different mass-accretion rates
         at different $T_{\rm eff}$s. }
\label{fig:acte}
\end{figure}

Figure \ref{fig:acte} gives the evolution of [Ca/He] on the 0.6M$_\odot$ DB WD
with different $\dot{M}_{\rm a}$ at different $T_{\rm eff}$s.
The timescale of reaching accretion-diffusion equilibrium is about $10^4$ for all models.
The mass-accretion rate
and the effective temperature greatly affect the element abundances of accreting
DB WD. When $\dot{M}_{\rm a}$ decreases from $10^{10}$ to $10^{4}$ g s$^{-1}$,
[Ca/He] reduces from about $10^{-5}$ to $10^{-10}$. It means that
the metal abundance of accreting WD is approximately in proportion to the mass-accretion rate.
In fact, \cite{Dupuis1992} and \cite{Koester2009} assumed that the element abundances observed in polluted WDs
should be accretion-diffusion equilibrium, and they suggested that the mass fraction of the i-th element ($X_{\rm cvz}$)
in the convective zone is given by
\begin{equation}
M_{\rm cvz}\frac{{\rm d}X_{\rm cvz, i}}{{\rm d}t}=\dot{M}_{\rm i}-\frac{X_{\rm cvz, i}M_{\rm cvz}}{\tau_{\rm diff, i}},
\label{eq:M}
\end{equation}
where  $X_{\rm cvz, i}$ and $\dot{M}_{\rm i}$ is the i-th element abundance in convective zone and the mass-accretion rate of i-th element.
Here, $\tau_{\rm diff,i}$ is the i-th element diffusive timescale, which can be estimated by
\begin{equation}
\tau_{\rm diff,i}=\frac{M_{\rm cvz}}{4\pi R_{\rm B, cvz}^2 \rho_{\rm B, cvz} v_{\rm diff, i}},
\label{eq:t}
\end{equation} where $R_{\rm B, cvz}$ and $\rho_{\rm B, cvz}$ are the radius and the mass density at the bottom of
the convective zone, respectively. Here, $v_{\rm diff, i}$ is the i-th element velocity of downward sedimentation at bottom of the convective envelope.
If $\tau_{\rm diff}$ is very shorter than WD lifetime, \cite{Koester2009} gave the relation between mass-accretion rate and the element abundance
by
\begin{equation}
X_{\rm cvz, i}=\frac{\dot{M}_{\rm i}}{M_{\rm cvz}}\tau_{\rm diff,i}.
\label{eq:x}
\end{equation}
Obviously, our result is consistent with Eq.(\ref{eq:x}).

However, the change of [Ca/He] with $T_{\rm eff}$ is complex.
When $T_{\rm eff}$ decreases from 20 to 10 kK,  [Ca/He] reduces by about three orders of magnitude.
When it decreases from 10 to 8 kK, [Ca/He] slightly enhances.
When it decreases from 8 to 6 kK, [Ca/He] reduces by about one order of magnitude again.
This change can be explained by the relation of $T_{\rm eff}$ and $M_{\rm cvz}$(See Figure \ref{fig:sconvt}).

\begin{figure}
\includegraphics[totalheight=3.5in,width=3.0in,angle=-90]{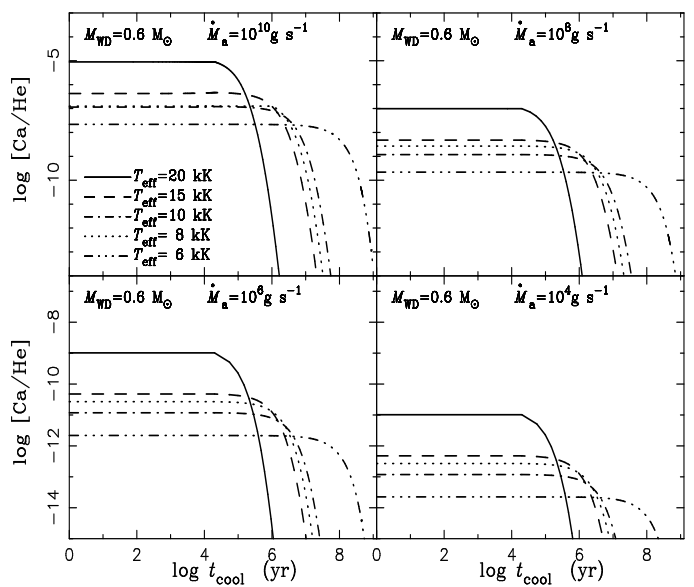}
\caption{Similar with Figure \ref{fig:acte} but for the evolution of [Ca/He] on the surface of 0.6 $M_\odot$ DB WD
         which just experiences lasting $10^6$ yr accretion at different $T_{\rm eff}$.
         The mass-accretion rates are showed in the middle-top region of
          every panel. }
\label{fig:qchtc}
\end{figure}

Figure \ref{fig:qchtc} gives the diffusion downward of Ca element on the surface of 0.6 M$_\odot$ DB WD
after a lasting $10^6$ yr accretion at different $T_{\rm eff}$.
The evolution of [Ca/He] with $T_{\rm eff}$ is similar with that in Figure \ref{fig:acte}.
In fact, Figure \ref{fig:qchtc} indicates the timescale of Ca element diffusion, that is $\tau_{\rm diff, Ca}$.
Obviously, it deeply depends on $T_{\rm eff}$. In our model, $\tau_{\rm diff, Ca}$ increases from about $10^5$
to $10^9$ yr when $T_{\rm eff}$ decreases from 20 to 6 kK. However, $\tau_{\rm diff, Ca}$ in \cite{Koester2009}
increases from about $10^4$ to $10^6$ yr.

\begin{figure*}
\includegraphics[totalheight=7.0in,width=3.5in,angle=-90]{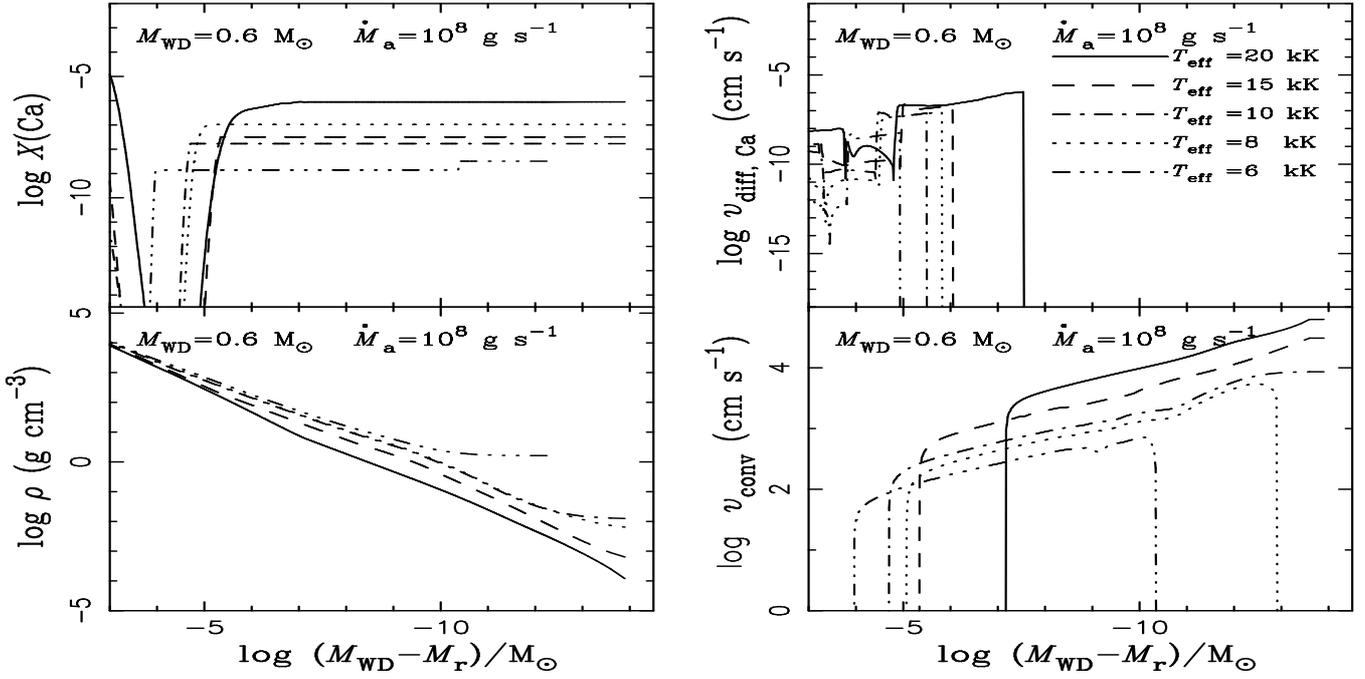}
\caption{The profiles of Ca abundance ($X{\rm (Ca)}$), opacity ($\kappa$), Ca diffusive velocity ($v_{\rm diff, Ca}$),
convective velocity ($v_{\rm conv}$), mass density ($\rho$) and radius ($R$)
around the surface of 0.6 $M_\odot$ DB WDs with
a mass-accretion rates of $10^8$ g s$^{-1}$ but
different effective temperatures which are represented by different lines.}
\label{fig:tvdif}
\end{figure*}

Based on Eq. (\ref{eq:t}), $\tau_{\rm diff, Ca}$ depends on $M_{\rm cvz}$,
$R_{\rm B, cvz}$, $\rho_{\rm B, cvz}$ and $v_{\rm diff, Ca}$.
Figure \ref{fig:tvdif} shows the profiles of $X{\rm (Ca)}$,  $v_{\rm diff, Ca}$, opacity($\kappa$) and $v_{\rm conv}$
around the surface of 0.6 M$_\odot$ DB WDs with different $\dot{M}_{\rm a}$ and and $T_{\rm eff}$.
Obviously, compared model of $10^8$ with  that of $10^6$ g s$^{-1}$,
the mass-accretion rate can affect $X{\rm (Ca)}$, but does not change the internal structure of
the accreting DB WD, including the opacity, the mass density and the radius.
The reason is that the matter accreted by DB WD quickly diffuses whole convective zone.
DB WD structure mainly depends on cooling duration which is presented by the effective temperature.
Due to massive convective zone of DB WDs, compared with He element, abundances of other heavy elements are very low.
Therefore, the accreted matter can not affect the internal structure of DB WD, which
depends on the cooling duration presented by $T_{\rm eff}$.

\begin{figure*}
\includegraphics[totalheight=7.0in,width=3.5in,angle=-90]{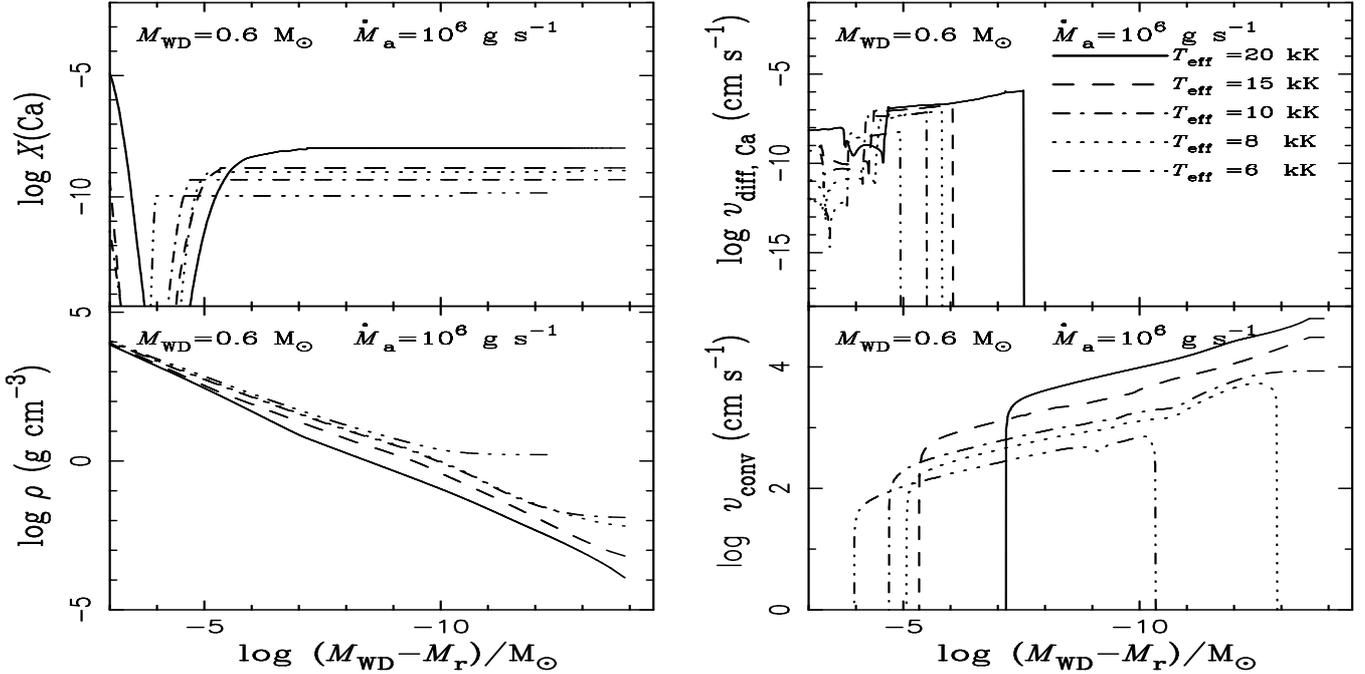}
\caption{Similar with Figure \ref{fig:tvdif} but for $\dot{M}_{\rm}=10^6$ g s$^{-1}$.}
\label{fig:tvdif7}
\end{figure*}

Combining Figures \ref{fig:sconvt}, \ref{fig:tvdif} and \ref{fig:tvdif7}, with DB WD cooling from $T_{\rm eff} = 20$ kK to 6 kK,
$M_{\rm cvz}$ increases from about $10^{-8}$ to $10^{-5}$ M$_\odot$,
$\rho_{\rm B, cvz}$ also increases from $\sim$ 1 to $10^3$ g cm$^{-3}$, $v_{\rm diff, Ca}$  at the base of the surface convection zone
decreases from about $10^{-6}$ to $10^{-10}$ cm s$^{-1}$, while $R_{\rm B, cvz}$ keeps constant. Therefore,
$\tau_{\rm diff, Ca}$ increases from  about $10^5$ to  $10^9$ yr. It means that, compared with $t_{\rm cool}$, $\tau_{\rm diff, Ca}$ can not be
neglected when $T_{\rm eff}<10$ kK, that is, Eq. (\ref{eq:x}) is not suitable for cool polluted DB WDs.
Of course, one should note, in our model, $v_{\rm diff, Ca}$ has irregular oscillations when $T_{\rm eff} < 8$ kK.
\cite{Koester2009} did not show $v_{\rm diff, Ca}$. However, $v_{\rm dif, Ca}$ may result
in great difference of $\tau_{\rm diff, Ca}$ between the present paper and \cite{Koester2009}.

\subsection{Accretion Pollution with Power Law}
\begin{figure}
\includegraphics[totalheight=3.5in,width=3.0in,angle=-90]{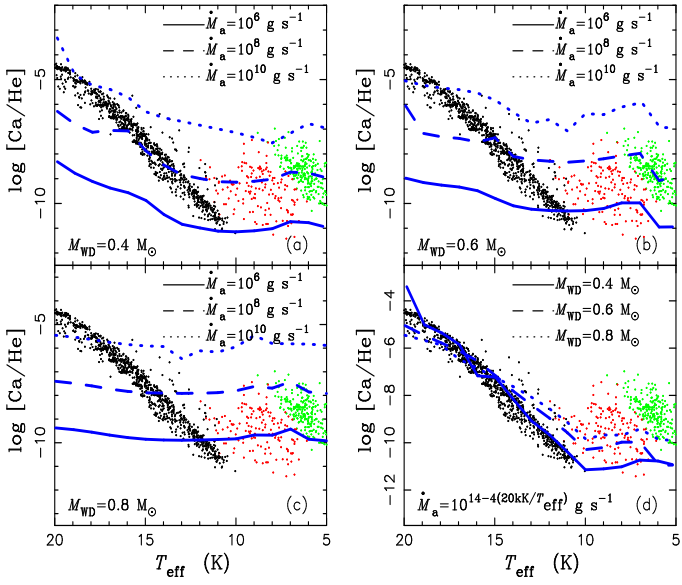}
\caption{Effective temperature vs. log [Ca/He] for DB WDs.
        Panels (a), (b), and (c) represent models with different mass DB WDs and
        a constant accretion rate ($10^{6}$, $10^{8}$ and $10^{10}$ g s$^{-1}$), respectively.
        Panels (d) is for the mass-accretion rate given by Eq. (\ref{eq:dotma}).
        Black, red and green dots represent observations from \cite{Koester2015}, \cite{Dufour2007} and \cite{Hollands2017},
        respectively. }
\label{fig:mact}
\end{figure}

As Figure \ref{fig:chte} shows, the [Ca/He] of about 1500 DB WDs observed
by \cite{Dufour2007},  \cite{Koester2015} and \cite{Hollands2017} must be explained
by accretion pollution.
In the panles (a), (b) and (c) of Figure \ref{fig:mact}, we give the evolutional tracks of [Ca/He] with $T_{\rm eff}$
for DB WDs with masses of 0.4, 0.6 and 0.8 M$_\odot$ and  constant mass-accretion rates of
$10^{10}$, $10^8$ and $10^6$ g s$^{-1}$, respectively.
Obviously, the results hardly explain the observations.


In order to model the metal abundances in GD 362's  atmosphere,
\cite{Jura2009} provided that $\dot{M}_{\rm a}$ decreases by power law:
\begin{equation}
\dot{M}_{\rm a}=\frac{M_{\rm disk}}{t_{\rm disk}}e^{-t/t_{\rm disk}},
\label{eq:dotm}
\end{equation}
where $M_{\rm disk}$ is the mass of planet disrupted by WD and $t_{\rm disk}$ is a characteristic timescale of accretion
disk. \cite{Jura2009} found that all of GD 362's distinctive properties can be explained if $M_{\rm disk}$ is between about
$10^{25}$ and $10^{28}$ g, in which the range of $t_{\rm disk}$ is between about $2\times10^5$ and $10^9$ yr.

However, based on Figures \ref{fig:acte} and panels (a) - (c) of Figure \ref{fig:mact},
with DB WD cooling, a decreasing mass-accretion rate with power law results in a continued decrease of
[Ca/He]. Therefore, it can not
explain the observations. According to our model, the mass-accretion rate should decrease when $T_{\rm eff}>$ 10 kK, but it should increase when $T_{\rm eff}<$ 10 kK.
Considering that $T_{\rm eff}$ of WDs mainly depends on $t_{\rm cool}$ and can compare with the observations, we assume that
$\dot{M}_{\rm a}$ changes by power law:
\begin{equation}
\dot{M}_{\rm a}=\left\{
\begin{array}{ll}
10^{14}\times10^{-4(\frac{20\ {\rm kK}}{T_{\rm eff}})}, {\rm\ g\ s^{-1}} & T_{\rm eff}>10 {\rm kK}\\
10^{3}\times10^{3(\frac{10\ {\rm kK}}{T_{\rm eff}})},\ \ {\rm\ g\ s^{-1}} & T_{\rm eff}<10 {\rm kK} \\
\end{array}
\right.
\label{eq:dotma}
\end{equation}

The panel (d) of Figure \ref{fig:mact} gives the evolution of [Ca/He] with $T_{\rm eff}$ for
DB WDs with different $M_{\rm WD}$ and an power-law $\dot{M}_{\rm a}$ described by Eq. (\ref{eq:dotma}).
Our results are consistent with the observations for DB WDs.
The $t_{\rm disk}$ of an accretion disk composed purely of dust is higher than $10^9$ yr\citep{Farihi2008}.
Usually, the cooling timescale of DB WD from 20 kK
to 10 kK is about $10^8-10^9$ yr, and it is about $10^9$ yr from 10 kK to 5 kK.
It means that a DB WD can accrete a disk produced by itself disrupting a planet during the whole cooling phase.
The decrease of mass-accretion rate when $T_{\rm eff}>10$ kK results from the viscous dissipation of accretion disk\citep{Jura2009}.
However, we do not find any model to explain its enhance when $T_{\rm eff}<10$ kK.
If Eq. (\ref{eq:dotma}) basically represents the true trend of the mass-accretion rates,
this indicates that the accretion disk produced by WD disrupting a planet may have complex structure.

\section{Conclusions}
In order to explain the evolutionary sequence of heavy element abundances observed among 1500 DB or DZ WDs,
we use MESA to create DB WDs with masses of 0.4, 0.6 and 0.8 M$_\odot$ by artificially stripping envelope once.
The H-rich envelope is stripped when stars evolve into red giants.
We investigate the effects of input parameters ($\alpha_{\rm MLT}$, $\alpha_{\rm th}$ and $Z$) on DB WD structures.
Due to the small pressure scale height, thick convective zone or mean molecular weight of DB WDs,
these input parameters have weak effect on DB WD structures including interior temperatures, chemical profiles and convective zones.
Therefore, they hardly affect the evolution of heavy elements on the surface of DB WDs.

Due to high gravitational fields of DB WDs, the element diffusion in the
theoretical model is too fast to explain the observations.
Therefore, the heavy elements on the DB WDs' surfaces
may originate from the pollution by accreting the planet disrupted by these WDs.
They mainly depend on the mass-accretion rates and the effective temperatures of
DB WDs.
In our model, a constant mass-accretion rate can not explain the evolutionary sequence
of Ca element for about 1500 observed DB or DZ WDs.
However, it is consistent well with the model in which the mass-accretion rate decreases
by one power law when $T_{\rm eff}>10$ kK and  slightly increases by another power law when $T_{\rm eff}<10$ kK.
The observed DB WD evolutionary sequence of heavy element abundances originates from WD cooling
   and the change of mass-accretion rate.

\begin{acknowledgements}
This work received the generous support of the National Natural Science Foundation of China,
project Nos. 11763007, U2031204, and 11863005.
\end{acknowledgements}

%
%

\end{document}